\documentstyle[psfig,twocolumn,prl,aps,amssymb]{revtex}
\newcommand{\be}{\begin{equation}}
\newcommand{\ee}{\end{equation}}
\newcommand{\bea}{\begin{eqnarray}}
\newcommand{\eea}{\end{eqnarray}}

\begin{document}
\wideabs{
\draft

\date{April 15, 2002} 
\title{Structures for Interacting Composite Fermions: Stripes,
Bubbles, and
Fractional Quantum Hall Effect}
\author{Seung-Yeop Lee, Vito W. Scarola, and J.K. Jain}
\address{Department of Physics, 104 Davey Laboratory, The
Pennsylvania State 
University, University Park, Pennsylvania 16802}
\maketitle

\begin{abstract}

Much of the present day qualitative phenomenology of the
fractional quantum Hall effect can be 
understood by neglecting the interactions between composite
fermions altogether. For 
example the fractional quantum Hall effect at $\nu=n/(2pn\pm 1)$
corresponds 
to filled composite-fermion Landau levels,  and the compressible
state at $\nu=1/2p$ to 
the Fermi sea of composite fermions.  
Away from these filling factors, the residual interactions
between composite 
fermions will determine the nature of the ground state. 
In this article, a model is constructed for the residual
interaction between 
composite fermions, and various possible states are considered
in a variational 
approach.  Our study suggests formation of composite-fermion
stripes, bubble crystals, 
as well as fractional quantum Hall states for appropriate
situations.

\end{abstract}

\pacs{71.10.Pm,73.43.-f}}


\section{Introduction}

In two dimensions, electrons subjected to a strong magnetic
field 
avoid one another by capturing an even number of 
quantized vortices and turning into composite fermions
\cite{Jain,review}.
The composite fermions interact much more weakly than electrons. 
In fact, 
tremendous progress can be made towards understanding the
dramatic behavior 
of this system by treating composite fermions as
non-interacting.
It is not that there is no interaction between the composite
fermions, but it  
is weak and often does not alter the nature of the state in a
qualitative manner, i.e., can be treated perturbatively.  In
such situations, while the 
inter-composite fermion (CF) interaction is surely important for
quantitative 
considerations, it does not affect 
the qualitative phenomenology, and can be neglected altogether
when the aim is to 
describe the qualitative phenomenology.

For example, the odd-denominator fractions $\nu=n/(2pn\pm 1)$
map into $\nu^*=n$ of composite
fermions carrying $2p$ vortices.  Here, the system has a gap
even in the absence of inter-CF 
interactions -- in stark contrast to the
{\em electron} system at $\nu=n/(2pn\pm 1)$ which had an
enormous degeneracy in the absence of
interactions -- and it is possible that for many of these
filling factors, the gap would not 
disappear as the inter-CF 
interaction is slowly turned on to its physical value.  If that
were the case, the
inter-CF interaction is unimportant at a qualitative level. 
This explains the fractional
quantum Hall effect \cite {Tsui} (FQHE) at filling
factors $\nu=n/(2pn\pm 1)$ in terms of the integral QHE \cite
{Klitzing}
of composite fermions \cite{Jain}.

Another example is at fractions $\nu=1/2p$, which map into
composite fermions at zero 
effective magnetic field.  If the interactions between composite
fermion are neglected, 
a Fermi sea is obtained \cite{HLR}.  
Conceptually, it is more difficult to justify the neglect 
of inter-CF interactions in this case, as there is no gap for
the Fermi sea.  Nonetheless, 
at least there is a well defined ground state for composite
fermions, 
unlike for non-interacting electrons at $\nu=1/2p$, and 
various experimental studies \cite{half} have shown that 
the CF Fermi sea is a good starting point for many purposes.

We recall that at relatively high temperatures and low
mobilities only the integral 
quantum Hall effect was seen, which is explained by neglecting
the interaction between 
electrons.  As the sample quality improved and temperature
lowered, FQHE was observed, 
which is understood in terms of weakly interacting composite
fermions.  
While the model in which the composite fermions are taken as
non-interacting has been
strikingly successful,  one may ask if another set of new
structures would emerge 
as the experiments further improve.
This question has motivated our investigation of the subtle
physics arising from 
inter-CF interactions.  Fortunately, there already exists a good
microscopic approach for  
investigating this issue, in the form of the wave functions for
composite fermions \cite{Jain}.  
Even though these wave functions are motivated by the physics of
independent composite
fermions, 
they give an excellent description of the interaction between
the composite fermions
\cite{Dev,JK,review}.  In fact, 
the wave functions practically give the exact solution to the
problem, 
which implies that they incorporate the full interaction
effects.  For example, consider the
low-energy branch of excitations, described as an exciton of
composite fermions.  If the
composite fermions were non-interacting, this branch would not
have any dispersion.  In reality,
there are oscillations in the energy as a function of the wave
vector, arising from the residual
interaction between the CF-particle and the CF-hole.  The
dispersion computed from the wave
functions provides an accurate account of the oscillations.  The
wave functions similarly obtain
the energy of two CF particles or two CF holes quite accurately. 

The inter-CF interactions are of course always relevant to
quantitative issues.  
However, there are two situations in which they can also make a
qualitative difference 
in the physics.  The first situation is when the
inter-CF interaction overwhelms the gap of the non-interacting
CF model, 
thereby destroying the FQHE\cite{Kamilla,Scarola2}.  It has been
shown that 
for $\nu\leq 1/9$ as well as for most
fractions in higher Landau levels, 
the FQHE undergoes an excitonic instability, i.e., the energy of
the
exciton becomes smaller than the energy of the uniform filled
CF-Landau level (LL) state.  The 
CF theory thus
not only tells us where FQHE occurs but also where it does not. 
(Recent 
experiments \cite {Pan} have shown evidence for FQHE at 1/9 at
{\em finite} temperatures.  
Even though this observation is not presently understood, it
ought to be stressed that 
it is not inconsistent with the earlier theoretical predictions
which only address the 
zero temperature phase.)  A pairing instability has also
been suggested at $\nu=5/2$, i.e., at the half filled second LL,
where the residual 
interaction  between the composite fermions is attractive
\cite{Scarola,Moore}.  The resulting 
instability of the CF Fermi sea causes a gap to open up here,
producing a plausible scenario 
for the 5/2 fractional quantum Hall effect \cite{Willett}.

The other situation where the inter-CF interaction is expected
to be important is away from the
special filling fractions $\nu=n/(2pn\pm 1)$, where the topmost
CF-LL is only partially occupied,
and the ground state has a large degeneracy if the inter-CF
interaction were to be turned off.
The degeneracy for independent composite fermions is much
smaller than for independent electrons 
even here, making the former a better defined starting point.   
The weak residual interaction between the composite fermions
will decide the nature of the 
true ground state here.  The present article deals with such
situations. 

The first possibility that comes to mind when composite fermions
partially occupy a 
Landau level is that 
of their {\em fractional} QHE.  This could happen when the
composite fermion 
filling factor is $\nu^*=n\pm \bar{n}/(2\bar{p}\bar{n}\pm 1)$,
which correspond, as usual, 
to electrons filling factors given by $\nu=\nu^*/(2p\nu^*\pm
1)$. 
We will consider the special case when the base 
composite fermions are carrying two vortices, denoted by
$^2$CFs, and the $^2$CFs 
in the topmost partially occupied CF-LL 
capture $2$ {\em additional} vortices to become $^4$CFs
(composite fermions carrying 
four vortices) and condense into an incompressible state with
$\bar{n}$ filled Landau 
levels.
The full state is then made up of two flavors of composite
fermions, with $^2$CFs in the lower 
filled CF LLs, and $^4$CFs in the topmost CF-LL.  Such states
are quite analogous to the 
FQHE state at $\nu=2+1/3$, where electrons in 
the second LL convert into composite fermions whereas the
electrons in 
the lowest LL remain electrons.  Such states have been
considered in the past \cite{Jain,Park2,Scarola3}, 
dubbed mixed flavor states of composite fermions. 
Of course, depending on the filling, it is also possible that
composite fermions would 
capture additional vortices to form a Fermi sea or a paired
state.  
Yet another possibility is for them to not capture any new
vortices but rather to phase 
separate in some manner, possibly through formation of stripes. 
Our aim in 
this article is to consider these possibilities as
quantitatively as 
presently possible over a range of filling factors.

It should be emphasized that while these states are easily
viewed in terms of 
composite fermions, they
are fantastically complicated in terms of electrons.  Consider
the possibility of 
pairing at $\nu=3/8$, for example \cite{Scarola3}. 
First, all electrons capture vortices to turn into $^2$CFs at
$\nu^*=1+1/2$; then,
those in the half filled CF-LL put on two more vortices, and
attempt to make a 
$^4$CF Fermi sea;  but the 
Fermi sea is unstable to pairing; pairing opens up a gap to
produce fractional 
quantum Hall effect.  Such a state could perhaps not be
envisioned  
without the knowledge of composite fermions.

In an earlier short article \cite{Seung}, we had studied
composite fermion 
fillings given by $\nu^*=n+1/2$ by this method.  The present
article gives further details 
and also considers a broader range of filling factors.  The plan
of the article 
is as follows.  In the next section, we enumerate the various
approximations that 
go into the calculation.  Sect. III discusses how we obtain the
pseudopotentials for 
the inter-CF interaction, and how we obtain from it a real space
interaction 
between composite fermions.  Certain preliminary observations
are made in Section IV.
The variational wave functions are given in Section V, and
Section VI discusses the 
results of our calculations.  In Section VII, we briefly
investigate the effect of 
finite thickness on our results.

\section{Assumptions}

We will be studying composite fermions at fillings of the type
$\nu^*=n+\nu_n^*$, where $n$ is
the number of filled composite fermion Landau levels and
$\nu_n^*$ is the filling 
fraction of the
topmost composite fermion Landau level.  In order to make
progress, we will make 
several assumptions.

The system will be taken to be fully spin polarized, which is a
valid approximation 
for sufficiently high magnetic fields.  The method can be
adapted to situations for which the 
states are only partially spin polarized,\cite{Scarola3} 
but here we will confine our attention to fully spin polarized
systems. 

We will map below the problem of interacting electrons into 
that of interacting composite fermions only of the topmost
partially filled CF Landau 
level.  The interaction between composite fermions will be
obtained by ``integrating 
out" the composite fermions of the lower filled CF-LLs.

We will assume that the filled CF Landau levels can be treated
as inert and
the problem can be defined solely in terms of the composite
fermions in the topmost
partially filled CF Landau levels.  This should be qualitatively
correct between well defined 
FQHE states described by filled CF Landau levels.  It should
also be a quantitatively 
accurate approximation when  
the CF-LL separation is large compared to the inter-CF
interaction.
Indeed, for filling factor in the range $1/3 < \nu < 2/5$ the
composite fermion theory
neglecting CF-LL mixing provides a very good description of the
actual states obtained from
exact diagonalization \cite{Dev,JK}.  
However, this approximation may become worse when the effective 
cyclotron energy is not too large; here  
the hybridization of CF states with higher CF Landau levels may
become relevant. 

Given that the composite fermion system actually represents a
strongly correlated state
of matter, the interaction energy for composite fermions in
general involves two-, 
three-, and $n$-body terms.  
The most critical assumption will be our neglect of three and
higher body terms.  
Following Wojs and Quinn \cite{Wojs} we will assume that the
interaction
energy of the many composite fermion system can be modeled as a
sum of pair-wise 
interaction.  Furthermore, we will determine the pair
interaction from the system 
that contains only two composite fermions in the partially
occupied CF Landau level 
(with the lower CF Landau levels fully occupied).  
This assumption is valid only if the
three and higher body terms do not cause a phase transition.  

There are indications that the higher order terms are small.
The {\em exact} energy spectrum of the system with three 
composite fermions in the second CF-LL (where the only input is
the Coulomb interaction
between the electrons) has been compared with the energy
spectrum that the pair-wise 
interaction predicts, and a reasonably good agreement is
obtained \cite{Wojs}.
The latter spectrum is obtained by diagonalizing the effective
two-body interaction
between composite fermions obtained by the method outlined
earlier.
However, no systematic study of the effects of 
multi-particle interaction has been performed as a function of
the number of particles.

Another limitation of our work is its variational nature, which
makes it as good as 
the trial wave functions for various states.  For example, 
for FQHE states, we will use the standard wave functions for
composite fermions \cite{Jain}.
These are extremely accurate in the lowest Landau level, but not
so accurate for FQHE in 
higher Landau levels.  Of course, it is also possible that the
actual ground state 
has some completely new structure, not described physically by
any of the trial 
states considered here.

These caveats are meant to stress right at the outset that the
results below are not 
``hard" predictions of the composite fermion theory, but are
predicated 
upon several assumptions.  This study is the first step
toward a theoretical treatment of the interaction between
composite fermions, and it 
is hoped that future experimental and theoretical studies will
shed light on 
the reliability of our present model.

\section{Model for Interaction between composite fermions}

\subsection{Pseudopotentials for CF-CF interaction}

It is known that a two-body interaction of fermions confined to
some Landau level 
is fully characterized  
by its Haldane pseudopotentials \cite{Haldane}, $V_m$, 
which are the energies of two particles in the 
relative angular momentum state $m$, $|\psi_m\rangle$: 
\bea
 V_m&=&\frac{\langle \psi_m|V|\psi_m\rangle }{ \langle
\psi_m|\psi_m\rangle}\\
&=&\frac{1}{2^{2m+1}m!\l_0^{2m+2}}\int
r\,dr\,V(r)\,r^{2m}e^{-r^2/4\l_0^2} \nonumber \\
&=&\int q\,dq\,\tilde V(q)e^{-q^2\l_0^2}L_m(q^2\l_0^2)
\nonumber \eea
Here $\l_0$ is the magnetic length, $L_m$ is a Laguerre
polynomial 
and $|\psi_m\rangle$ is the two-body state of relative angular
momentum $m$. 
For fully spin polarized fermions, only the pseudopotentials
$V_m$ with 
odd integer values of $m$ are relevant.
In this paper, the Fourier transformation is defined
symmetrically
\bea\tilde V(q)\equiv\frac{1}{2\pi}\int d^2r\,
V(r)e^{-i\vec q\cdot\vec r}=\int r\,dr\,J_0(qr) V(r)\eea 
where $J_0$ is a Bessel function. 
In the equations below, the interelectron interaction, 
$V(r)=e^2/\epsilon r$, will be expressed in units of
$e^2/\epsilon \l_0$, and $r$ in 
units of $\l_0$;  the interelectron interaction then becomes
$V(r)=1/r$.  The 
appropriate units will be put back at the end. 

Given the pseudoptentials 
we can construct the two body interaction in the lowest LL using 
the following formula.
\bea \tilde V(q)=2\pi\sum_m V_m L_m(q^2).\eea
Alternatively, we can directly write the Hamiltonian as  
\bea H|\psi\rangle=\sum_{i<j}\sum_{m}V_mP_m^{ij}|\psi\rangle
=\sum_{i<j}\sum_{m= odd}V_mP_m^{ij}|\psi\rangle.
\eea
Here $P^{ij}_m$ projects particles $i$ and $j$ onto a state of 
relative angular momentum $m$. 
This Hamiltonian is useful for exact diagonalization study. 

The pesudopotentials for composite fermions are defined
similarly.  For 
the pseudopotentials in the $n$th composite fermion Landau
level, 
we consider the state in which the lower CF Landau levels are
fully 
occupied and the $(n+1)^{st}$ CF Landau levels contains two
composite fermions.  
$V^{CF}_m$ is then defined as the energy of the state in which
the 
two composite fermions are in the relative angular momentum $m$
state.
In other words,
\bea 
V^{CF}_m=\frac{\langle\psi^{CF}_m|\sum_{i<j}\frac{1}{r_{ij}}
|\psi^{CF}_m\rangle}{\langle\psi^{CF}_m|\psi^{CF}_m\rangle}-E_{\infty}
\eea
where $|\psi^{CF}_m\rangle$ is the wave function of composite
fermions in 
the relative angular momentum $m$ states.  Note that the state
of two 
composite fermions in the $(n+1)$st level 
is a many body state of electrons. For convenience, the 
energy is measured relative to $E_{\infty}$, the energy of the
state in which the 
two composite fermions are far separated.  (For finite systems
the value of 
$E_{\infty}$ is chosen in the manner explained below.)

We use the spherical geometry for our
calculations.\cite{Haldane} 
Following the standard procedure for writing the wave functions
for 
composite fermions \cite{Jain} the wave function for the
relevant state with angular 
momentum $L$ is given by 
\begin{equation}
|\psi^{CF}_L\rangle={\cal P}_{LLL}\Phi^2_1\Phi^{++L}_n, 
\end{equation}
where $\Phi_1$ is the wave function of one filled LL,
$\Phi_n^{++L}$ 
is the wave function of the electron state in which $n$ LLs are
fully 
occupied and the $(n+1)$st LL contains two electrons in  
angular momentum $L$ state:
\bea
\Phi^{++L}_n=\sum_{m_1,m_2}
<LM|l m_1; l m_2> a^{+}_{n+1,m_1}a^{+}_{n+1,m_2}\Phi_n
\eea 
Here $\Phi_n$ is the state with n-filled LLs,  
$a^{+}_{n+1,m_1}$ creates an electron 
in the $l_z=m_1$ state in the $(n+1)$th LL,
$l$ is the
angular momentum of a single composite fermion in the
$(n+1)^{st}$ LL, 
and $<LM|l m_1; l m_2>$ are Clebsch-Gordon coefficients.
Because the energy depends only on the total angular momentum
$L$, we work 
with $M=0$, i.e., $m_1+m_2=0$.
The explicit form for the general wave functions of this kind 
is given in the literature \cite{JK}.
${\cal P}_{LLL}$ is the lowest LL projection operator.

In the spherical geometry, $V_m$ is defined 
as the energy of the state with angular momentum $L=2l-m$
\cite{Quinn}:
\bea 
V^{CF}_m=\frac{\langle\psi^{CF}_{2l-m}|\sum_{ij}\frac{1}{r_{i<j}}
|\psi^{CF}_{2l-m}\rangle}{\langle\psi^{CF}_{2l-m}|\psi^{CF}_{2l-m}\rangle}-E_{\infty}
\eea
This brings the definition of the pseudopotentials into
conformity with 
the disk geometry, reproducing the disk pseudopotentials in the 
thermodynamic limit $N\rightarrow \infty$, when the angular
momentum of the lowest 
Landau level shell also diverges: $l\rightarrow \infty$.  To see
this, it is helpful 
to note that in the disk geometry, 
the separation between the two particles increases with $m$.  In
the spherical 
geometry, the smallest separation is obtained when both
particles are put 
in the same $l_z$ state, which corresponds to the largest
angular momentum 
$L=2l$ (not bothering about the Pauli principle here).  Thus,
the 
largest $L$ corresponds to the smallest $m$.

The integrals are performed using Monte Carlo;  we evaluate
$V^{CF}_m$ 
for up to 50 particles. 
Of particular interest is the short range part of the
interaction, which appears to 
converge to the thermodynamic limit fairly quickly with the
number of particles, 
as expected.  The long range part of the interaction is
explicitly known, as 
discussed below.

In order to minimize the computation time, we calculate several
pseudopotentials 
within one Monte Carlo run by sampling with one wave function 
$\psi_{m_s}^{CF}({\bf r})$, according to 
\bea 
V_m^{CF}&\stackrel{*}{=}&\int d{\bf r} |\psi_m^{CF}({\bf r})|^2
\sum_{i<j} \frac{1}{r_{ij}}\\
&=&\int d{\bf r}\left(
\frac{ |\psi_m^{CF}({\bf r})|^2}{|\psi_{m_s}^{CF}({\bf r})|^2}  
\sum_{i<j} \frac{1}{r_{ij}}\right)  |\psi_{m_s}^{CF}({\bf r})|^2
\eea
where $(\stackrel{*}{=})$ means that the two sides are equal
upto an additive constant.
Generally, the error increases as the angular momentum moves
away from the sampling 
angular momentum, and one must eventually use a different state
for sampling; however,
sampling with a single state works satisfactory for $N\leq 40$,
where 
$N$ is the total number of particles.

In Monte Carlo with composite fermions, all elements of the
determinant are altered 
even if a single particle is moved, because of the strongly
correlated nature of the problem.
However, in a given step, the determinants that combine to give
the state with a well 
defined $L$ differ only in two columns.   Rather than
calculating each determinant entirely, 
we use an updating trick earlier used in Ref.~\cite{Park4},
which is a generalization of 
a method used for updating determinants that differ only in one
column \cite{Ceperley}. 
The method is explained in the Appendix.

The pseudopotentials for the CF-CF interaction are shown in
Fig.~1 of Ref.~\onlinecite{Seung}
 and Fig~(\ref{g2}). The explicit values are given in Tables
\ref{tab1}-\ref{tab6}.

\subsection{Inter-CF interaction in real space}

Now that the composite fermions of the fully occupied CF-LLs
have been 
integrated out to produce an effective interaction between the
composite fermions
of the partially filled CF-LL, only the $N$ composite fermions 
in the topmost partially filled LL will be explicitly considered
in what follows.  
(Earlier, the symbol 
$N$ was used for the {\em total} number of composite fermions,
but below it will 
be reserved for the number of composite fermions in the
partially filled CF-LL.) 

We thus have a system of fermions restricted to 
a certain CF LL, with an interaction known through its
pseudopotentials. 
It would be most natural to feed the pseudopotentials into an
exact diagonalization 
routine and compute the ground state.  However, such studies are
not useful for 
some of the states that we wish to consider.  The system sizes
accessible 
to exact diagonalization are too small to capture the physics of 
compressible or stripe states.  (Also, the spherical 
geometry often used for exact diagonalization is not
particularly convenient for 
the stripe phase.)

We shall instead proceed by considering variational wave
functions for 
several different kinds of states, and determining which has the
lowest energy.
As mentioned earlier, this approach has the shortcoming that we
do not know if we have missed the 
true ground state, and even for a given state, we may not have a
sufficiently accurate
wave function.  We will consider all of the states that have
been 
investigated in the past in the lowest and higher Landau levels
of {\em electrons}, and 
hope that the actual ground state is one of them.  Ultimately,
only 
experiments can tell us for sure if that is the case.

We will evaluate the energies of the some of the trial wave
functions using the Monte Carlo method, 
for which we will need a real space form for the interaction 
between composite fermions.  Furthermore, it is convenient to
write various 
trial wave functions in the lowest Landau level.  So, we will
map the problem of 
composite fermions at $\nu^*=n\pm \nu_n^*$ into 
that of $N$ fermions in the {\em lowest} LL at $\nu_n^*$.
Such a mapping can, in principle, be carried out exactly,
because the problem of 
fermions in any given LL with one interaction 
is equivalent to that of fermions in a different LL with another
interaction, provided 
the Haldane pseudopotentials for the two interactions are the
same.
Defining the problem in the lowest LL helps us to find 
the energies for various composite fermion states, because
convenient techniques exist for 
writing the wave functions in explicit form in the lowest LL.

For some of the states (paired state, Fermi sea) the calculation
will be done in 
the spherical geometry, and the energy in the thermodynamic
limit ($N\rightarrow \infty$) 
will be estimated by an extrapolation of finite system results. 
For the stripe and bubble phases, the calculations will be done
directly for the 
thermodynamic limit in the planar geometry.  In this case, the
spherical geometry 
is used only for the determination of the effective interaction
between composite fermions.

In order to map the problem of 
composite fermions at $\nu^*=n\pm \nu_n^*$ 
into fermions at $\nu_n^*$ in the lowest Landau level, we need
to 
find an effective interaction 
$V_{eff}(r)$ which produces the desired pseudopotentials in the
lowest Landau level.
A proper consideration of the long range part of the interaction
is important  
for our purposes below, especially in the context of stripes,
the period of which is 
determined by a competition between the short range attraction
and long range repulsion.
Therefore, we treat the long distance part of the interaction
explicitly in our calculations.
Fortunately, the form of the interaction in this limit is known:
it is 
given by the Coulomb interaction between particles of fractional
charge $e/(2n+1)$, i.e., 
\be\frac{(2n+1)^{-2}}{r}
\left[\frac{e^2}{\epsilon\l_0}\right]\ee
where $r$ is the dimensionless distance between the composite
fermions
measured in units of ${\l_0}$, the actual magnetic length at
$\nu$.
At the effective magnetic field, i.e., the magnetic field
corresponding to the filling factor 
$\nu^*$, the natural unit for length is $\l^*$.  We denote the
distance measured in units of $\l^*$ by $r^*$.  The long range
part 
then translates into:
\begin{equation}
\frac{(2n+1)^{-2}}{r^*}\frac{\l_0}{\l^*}\left[\frac{e^2}{\epsilon\l_0}\right]
\rightarrow
\frac{(2n+1)^{-5/2}}{r^*}\left[\frac{e^2}{\epsilon\l_0}\right]
\end{equation} 
where we have used $\l_0/l^*=(2n+1)^{-1/2}$.  
We model the full interaction between composite fermions at the
effective 
magnetic field $B^*$ by the following form:
\begin{equation}
V^{eff}(r^*)=\left(\sum_i c_i r^{*2i}e^{-r^{*2}}+ 
\frac{(2n+1)^{-5/2}}{r^*}\right)\left[\frac{e^2}{\epsilon
l_0}\right]
\end{equation}
The power series, with properly chosen coefficients $c_i$, 
takes care of the short range 
part of the inter-CF interaction.  The functions of the kind
$r^{*2i}e^{-r^{*2}}$ 
are used because the pseudopotentials for such functions are
analytically calculable. 
Many other functional choices could work equally well; the
mapping from  
pseudopotentials to a real space interaction in the lowest LL is
one-to-many, 
and all real space interactions with the same pseudopotentials
are completely equivalent in
the absence of LL mixing.  (Slightly different functions were
chosen in Ref.\cite{Park}.
There, the long rage Coulomb interaction was not considered
explicitly, 
because that work dealt with the composite fermion Fermi sea,
which corresponds 
to the limit $n\rightarrow \infty$, so the last terms vanishes.)
It is important to 
choose a real space form for the effective interaction
that makes the interaction as smooth as possible; 
large oscillations of the interaction lead to larger errors in
the Monte Carlo evaluation of the energy.

The units for the effective potential defined above ought to be
noted.  If we replace 
$e^2/\epsilon l_0$ by $e^2/\epsilon l^*$, then the interaction
has the natural form 
for a problem at the {\em effective} magnetic field.  We do our
calculations with this 
interaction potential, and then obtain the final energies by a
change of units.

Next, we proceed to find values for the coefficients 
$c_i$ so that the pseudopotentials of $V^{eff}$ reproduce the
pseudopotentials 
$V_m^{CF}$ calculated earlier from microscopic wave functions. 
Because we are considering fully polarized states, only the odd
pseudopotentials 
are relevant.  In principle, it is possible to fix all of the
$c_i$'s to reproduce all 
$V^{CF}_m$'s exactly, but in practice, we will fix the first
several (five to six) odd 
pseudopotentials only. That is sufficient for our purposes,
given many other approximations
that go into the calculation.  The energy is determined 
predominantly by the short distance part of the interaction.
Of course, at long distances, the above model interaction
automatically gives the correct behavior, thereby ensuring that
it  
is quite accurate almost everywhere.

The energy of two composite fermion particles evaluated from
microscopic wave functions
gets contribution from two sources, their self energies and
their interaction.  The self 
energy can be identified by putting them far apart; because 
it is constant, it does not affect the
calculation in any way.  We find it convenient to subtract it
out by shifting all  
$V_m^{CF}$ by a constant so that the last pseudopotential
(remember that there are only a
finite number of pseudopotentials in a finite system) becomes
equal to the pseudopotential of
the {\em long range} part of the model interaction:
\bea
E_{\infty} =
\frac{\langle\psi_{m_{\max}}|\frac{(2n+1)^{-\frac{5}{2}}}{r}|\psi_{m_{\max}}\rangle}
{\langle\psi_{m_{\max}}|\psi_{m_{\max}}\rangle}
\eea
where $|\psi_{m_{\max}}\rangle$ is the two-body state with
relative angular momentum $m$
in the LLL.  It is expected that for sufficiently large systems,
the contribution of the 
gaussian terms to the last pseudopotential is negligible due to
their rapid decrease
with distance.

We have matched the first five or six odd pseudopotentials 
by adjusting the same number of coefficients $c_i$.  The values
of $c_i$ used below are 
quoted in Tables (\ref{b1}) and (\ref{b2}).  
The comparison between the pseudopotentials $V_m^{CF}$,
calculated from wave 
functions, and the pseudopotentials calculated from the real
space model interaction
are shown in Fig.~1 of Ref.~\onlinecite{Seung} and \ref{g2},
indicating that the 
real space interaction provides a good representation of the
$V_m^{CF}$.
The specific choice of which $c_i$ are taken to be non-zero  
in our work (Tables \ref{b1} and \ref{b2}) is dictated by the
smoothness of the real space form 
of the interaction.  We also note that the form of
$V^{eff}(r^*)$ depends on details, 
and can appear rather strange, 
with oscillations and even attractive regions, but proper
results are guaranteed so long
as it generates the correct pseudopotentials. 

\section{Preliminary observations}

A glance at the CF pseudopotentials reveals one 
of the most striking features of the inter-CF interaction: it is
often 
attractive.  The smallest energy is obtained at the smallest 
relative angular momentum, i.e., at the shortest distance.
This is true for CF particles in the third or higher CF-Landau
levels
and for CF holes in the second and higher CF-LLs.
This fact already appears to suggest an answer to one important
question:  Are there any
new FQHE states between the fractions $n/(2n+1)$? New FQHE
states here can appear if $^2$CFs in 
a partially filled CF-LL capture additional vortices to form a
FQHE state.  However, 
capture of vortices requires a sufficiently strong short range
repulsion.  Given that  
the inter-CF interaction is actually attractive, any new FQHE in
the range $2/5<\nu<1/2$ appears
unlikely (for fully polarized systems), to the extent that our
model for interacting 
composite fermions is trustworthy.  
(The caveat is important; for example, three body interaction
could play an important role, or
when there are many composite fermions, the nature of the 
two-body interaction might change slightly from our model which
considers only two composite
fermion particles.)
Nonetheless, the form of the interaction suggests that new FQHE
states in the range $2/5<\nu<1/2$ 
are either very weak or altogether absent.  That is entirely
consistent with experiments.

\begin{figure}
\centerline{\psfig{figure=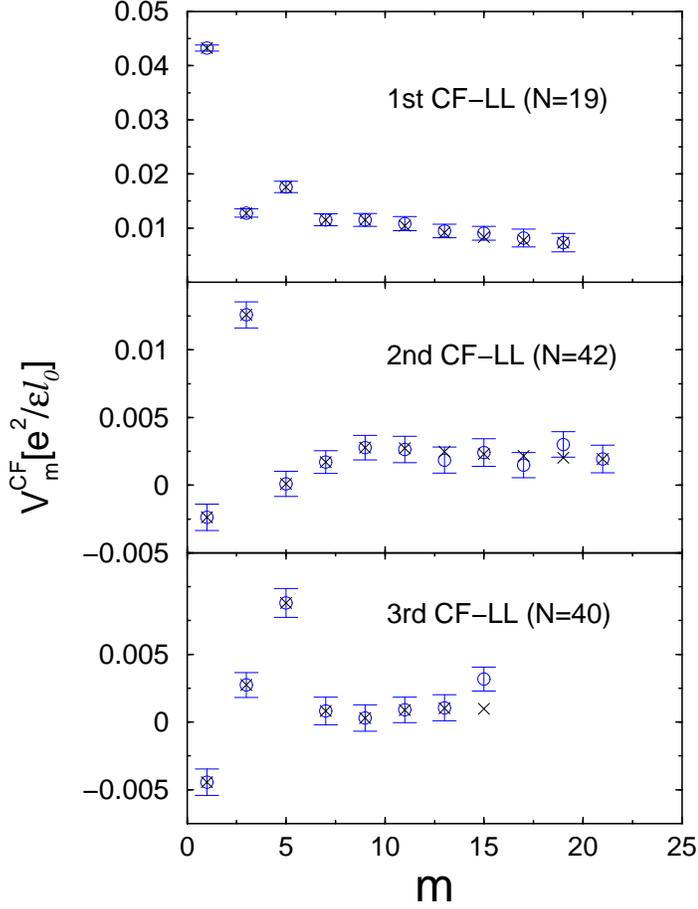,width=4.0in,angle=0}}
\caption{This figure gives the 
pseudopotentials for composite fermion ``holes" in the first, 
second, and third composite fermion levels.  These are defined
(up to an overall additive 
constant) as the energies of
two composite fermion holes in the appropriate composite fermion
Landau level, with the lower CF 
levels fully occupied. The systems are large enough that the
pseudopotentials at the first few $m$
(relative angular momentum) values are well converged. Circles
denote the pseudopotentials
computed from the microscopic wave functions, with error bars
showing the Monte Carlo uncertainty, and crosses
are the pseudopotentials for the real space model interaction
explained in the text.
The pseudopotentials for CF ``particles" in the second, third,
and fourth composite fermion
levels were given in Ref.~\protect\cite{Seung}.} 
\label{g2}
\end{figure}

For composite fermions in the lowest CF Landau level
($\nu<1/3$), on the other hand, 
the interaction is strongly repulsive (Fig.~\ref{g2}, top
panel), and new FQHE is expected.  That is also consistent with
experiment.
The $^2$CFs capture two additional flux quanta 
to become  $^4$CFs, which then fill Landau levels to produce
FQHE at $\nu=n/(4n\pm 1)$.
There is presently experimental evidence for six members of
these sequences \cite{Pan}.
These states can be interpreted  
either as the IQHE of $^4$CFs or as the FQHE of $^2$CFs; they
are 
``pure," (as opposed to ``mixed") in the sense that they contain
only a single 
flavor of composite fermions.

The CF pseudopotentials in Fig.~1 of Ref.~\onlinecite{Seung} and
Fig.~(\ref{g2})
of this paper also indicate that the assumption of treating the
interaction between
many composite fermions as the sum of two body interactions is
not unreasonable.  In a given CF-LL,
Fig.~1 of Ref.~\onlinecite{Seung}
 gives the interaction between two CF particles in an otherwise
empty CF-LL, whereas
Fig.~\ref{g2} gives the interaction between two CF holes in an
otherwise full CF-LL.  The latter
is a many body state of CF particles.  However, the two
interactions are rather similar both in
shape as well as magnitude, which is what would be expected if
only two body 
terms were relevant.

\section{Variational states}

\subsection{$^4$CF FQHE / $^4$CF Fermi sea}

One possibility for composite fermions in the partially filled
$^2$CF-LL 
is to capture two additional vortices to 
turn into $^4$CFs and then form an integral quantum Hall state
with $n'$ filled Landau 
levels, which corresponds to 
\begin{equation}
\nu^*=n\pm \frac{n'}{2n'\pm 1}
\end{equation}
filling for $^2$CFs, and to
\begin{equation}
\nu=\frac{\nu^*}{2\nu^*\pm 1}
\end{equation}
for electrons.

Taking $^2$CFs as base particles, 
the wave function for their FQHE state is given by:
\begin{eqnarray}
\Psi^{FQHE}_{\frac{n'}{2n'+ 1}}=P_{LLL} \Phi_{1}^{2p}\Phi_{n'}
\end{eqnarray}
where $\Phi_{n'}$ is the wave function of $n$ fully filled LLs
and $P_{LLL}$
projects the state onto the lowest LL.  Recall only the $^2$CFs
in the partially filled 
CF-LL are considered explicitly here; the filled $^2$CF Landau
levels have been 
integrated out in producing the effective interaction.
The state at $\nu^*=n+1/2$ is obtained in the limit
$n'\rightarrow \infty$, where 
the filled Landau level state $\Phi_{n'}$ becomes a Fermi sea.
We expect these states to be energetically 
favorable when $V_1^{CF}$ is large compared to other
pseudopotentials.

The energies of these states can be calculated with the help of
the real 
space interaction, following the method outlined in the
literature.
The spherical geometry will be used in our calculations, and the
thermodynamic 
energy will be obtained from a linear, least squares fit of the
finite system energies 
plotted as a function of $1/N$.

\subsection{$^4$CF pairing}

At half filled Landau level, pairing of composite fermions also
becomes possible.
A weak repulsion between fermions is believed 
to lead to pairing of composite fermions at half filling
\cite{Scarola,Moore,Scarola3,Park}.  
The second LL 
coulomb interaction between electrons falls within this class of 
interactions. The paired state is reasonably well described by a
Pfaffian wave function 
\cite{Moore}, which, in particular, has been shown to have lower
energy 
than the CF Fermi sea at $\nu=5/2$\cite{Park}.  Numerical
studies on small systems 
have shown that this state has a fairly large overlap with the
exact 
ground state \cite{Scarola3}, although the overlaps are not
decisive (in contract to the  
case for filled composite fermion Landau levels for which the
overlaps are in 
excess of 99\%).   

We ask whether or not the interaction between CFs at half filled 
CF LLs favors the paired state of $^2$CFs.  Here,  $^2$CFs
capture two 
vortices to turn into $^4$CFs and then form pairs.  With $^2$CFs
as base particles 
at filling factor equal to 1/2, we consider the Pfaffian wave
function for the paired 
state:
\begin{eqnarray}
\Psi^{Pf}_{\frac{1}{2}}=\Phi_{1}^{2}Pf\left[M\right]
\end{eqnarray} 
where $Pf\left[M\right]$ is the Pfaffian of the matrix $M$ with
components
$M_{ij}=(z_i-z_j)^{-1}$ defined as $Pf\left[M\right]\propto
A\left[
M_{12}M_{34}...\right]$.  $Pf\left[M\right]$ is a real space BCS
wave
function and so $\Psi^{Pf}_{\frac{1}{2}}$ can be viewed as a
p-wave 
paired quantum Hall state of $^4$CFs.  
We calculate the energy of this state as 
discussed in the previous section.

\subsection{$^2$CF stripes and bubble crystals}

For an attractive interaction, phase separation is a likely
possibility, 
which, from our experience with higher Landau level physics,
would lead 
to the formation of stripes or bubble crystals.  In this case,
the base 
particles, $^2$CF, do not capture any additional vortices.
We calculate the energies of these states closely following  
the Hartree-Fock formulation used earlier \cite{Fogler,Yoshioka}  
for electrons in higher Landau levels;  the only difference 
is the form of the interaction.  We give here a brief outline of 
the calculation for completeness, which closely follows
Ref.\onlinecite{Fogler}.

Consider fermions in the plane of size $L_x\times L_y$.
In the lowest LL the eigenstates are
\be
\psi_k(x,y)=\frac{1}{\sqrt{L_y l_0\pi^{1/2}}}e^{iky}
e^{-\frac{1}{2 l_0^2}(x+kl_0^2)^2}
\ee
where $k\in\frac{2\pi}{L_y}\times{\mbox{(integer)}}$.
Define the operator 
$\Psi({\bf r})\equiv\sum\nolimits_{k}\psi_k({\bf r})\,{a_k}$ 
where $a_k$ is the annihilation operator corresponding to the
state $|\psi_k>$.
With the Fourier transform defined as 
\bea\tilde f({\bf q})=\frac{1}{2\pi}\int d^2{\bf r}\,
f({\bf r})e^{-i{\bf q}\cdot{\bf r}}\eea
and 
\bea f({\bf r})=\frac{1}{2\pi}\int d^2{\bf q}\,
\tilde f({\bf q}) e^{i{\bf q}\cdot{\bf r}}\eea
we have
\bea f({\bf r})\approx\sum_{\bf q}\frac{2\pi}{L_xL_y}\tilde 
f({\bf q})e^{i{\bf q}\cdot{\bf r}}\eea
The Fourier transform of the density operator, 
$\rho({\bf r})\equiv\Psi^{\dagger}({\bf r})\Psi({\bf r})$, is
given by 
\bea\rho({\bf q})&=&\frac{1}{2\pi}\int d^2{\bf r}\,\rho({\bf
r})e^{-i{\bf q}\cdot{\bf r}} 
\nonumber \\
&=&\frac{1}{2\pi}\int d^2{\bf
r}\,\sum_{k,k'}\psi^{\dagger}_k({\bf r})\psi_{k'}({\bf r})
e^{-i{\bf q}\cdot{\bf r}}a^{\dagger}_ka_{k'}\eea
With the following change of variables
\bea
q_y=k'-k,~~~~~~~k_0=\frac{k+k'}{2},~~~~~~~k_{\pm}=k_0\mp
\frac{q_y}{2}
\eea
the density operator becomes
\bea
\rho({\bf q})&=&\frac{1}{2\pi^{3/2}l_0}\sum_{k_0}\int dx\times\\
&&e^{-\frac{1}{2l_0^2}(x+k_+l_0^2)^2-\frac{1}{2l_0^2}(x+k_-l_0^2)^2-iq_xx}
a^{\dagger}_{k_+}a_{k_-} \nonumber \\
&=&\frac{1}{2\pi}\sum_{k_0}\exp\left[-\frac{1}{4}q^2
l_0^2+ik_0q_x l_0^2\right]
a^{\dagger}_{k_+}a_{k_-}
\eea
Define the interaction operator ${\hat V}$:
\bea
{\hat V}&=&\frac{1}{2}\int d^2{\bf r}\,d^2{\bf r'}V({\bf r},{\bf
r'})\rho({\bf r})\rho({\bf r'})
\nonumber \\
&=&\frac{(2\pi)^3}{2L_xL_y}\sum_{\bf q}\tilde V({\bf
q})\rho(-{\bf q})\rho({\bf q})
\nonumber \\
&=&\frac{\pi}{L_xL_y}\sum_{\bf q}\sum_{k_0,k'_0}\tilde V({\bf
q})\times
\nonumber \\
&&\exp\left[-\frac{1}{2}q^2 l_0^2+
i(k'_0-k_0)q_x
l_0^2\right]a^{\dagger}_{k'_+}a_{k'_-}a^{\dagger}_{k_-}a_{k_+}
\eea
In the Hartree-Fock (HF) approximation, the energy of the system
is given by
\bea
\langle{\hat V}\rangle=\frac{\pi}{L_xL_y}
\sum_{{\bf q},k_0,k'_0}\tilde V({\bf q})e^{-\frac{1}{2}q^2
l_0^2+i(k'_0-k_0)q_x l_0^2}\times
\nonumber \\
\bigg[\langle a^{\dagger}_{k'_+}a_{k'_-}\rangle
\langle a^{\dagger}_{k_-}a_{k_+}\rangle-\langle 
a^{\dagger}_{k'_+}a_{k_+}\rangle\langle
a^{\dagger}_{k_-}a_{k'_-}\rangle\bigg]
\eea
In terms of the operator 
$\Delta({\bf q})\equiv\frac{1}{2\pi}\sum\nolimits_ke^{-ikq_x
l_0^2}\langle 
a^{\dagger}_{k_+}a_{k_-}\rangle$, which represents the density
of orbit centers, 
the energy is 
\bea
\langle {\hat V}\rangle&=&\frac{(2\pi)^3}{2L_xL_y}\sum_{\bf q}
[\tilde U(q)- l_0^2 U(q l_0^2)]\Delta(-{\bf q})\Delta({\bf q})
\nonumber \\
&=&\frac{1}{2}\int d^2{\bf r}\,d^2{\bf r'}\left[U(|{\bf r}-{\bf
r'}|)-
\frac{1}{l_0^2}\tilde U(\frac{|{\bf r}-{\bf
r'}|}{l_0^2})\right]\times \nonumber \\
&&~~~~~~~~~~~~~~~~~~~~~~~~~ \Delta({\bf r})\Delta({\bf r'})
\eea
where $\tilde U(q)\equiv V(q)e^{-\frac{1}{2}q^2 l_0^2}$ and
$U(r)$ is 
its Fourier transform. It is convenient to define 
$\tilde U_{HF}(q)=\tilde U(q)-l_0^2 U(q l_0^2)$ where the first
term 
corresponds to the direct interaction and the second to the 
exchange interaction.
$\langle{\hat V}\rangle$ in the above equation gives the CF-CF  
interaction energy, to which the CF-background and
background-background interaction
energies must be added to obtain a finite result. Following
Ref.~\ref{Fogler}, we
 define the cohesive energy as the 
energy difference measured from the {\em uniform} Hartree-Fock
state
\bea
E_{coh}&=&\frac{1}{N}\frac{(2\pi)^3}{2L_xL_y}\times
\nonumber \\
&&\sum_{{\bf q}\neq 0}[\tilde U(q)-l_0^2 U(q l_0^2)]\Delta(-{\bf
q})\Delta({\bf q})
\eea

We can now calculate the cohesive energies of the stripe and
bubble phases. 
In the stripe phase
with the periodic length of stripe $\Lambda$, 
the orbit-center density is written as 
\bea
\Delta(x,y)=\frac{1}{2\pi l_0^2}\sum_q
\frac{2\sin(\frac{q\Lambda\nu}{2})}{\Lambda q}e^{iqx}
\eea
where
$\{q\}=\frac{2\pi}{\Lambda}\times\{\cdots,-3,-2,-1,1,2,3,\cdots\}$.
So the cohesive energy is 
\bea
E_{coh}=\frac{1}{2\nu l_0^2}\sum_q\tilde U_{HF}(q)
\left(\frac{2\sin(\frac{q\Lambda\nu}{2})}{\Lambda q}\right)^2
\eea
In the bubble phase with the lattice constant $\Lambda_b$ for
the hexagonal lattice, radius of 
the bubble $R=\sqrt{\frac{\sqrt{3}\nu}{2\pi}}\Lambda_b$ and area
of a cell
$A=\frac{\sqrt{3}}{2}\Lambda_b^2$, the orbit-center density is 
\be
\Delta({\bf r})=\frac{1}{A}\sum_{\bf q}\frac{R}{l_0^2 q}
J_1(qR)e^{i{\bf q}\cdot{\bf r}}
\ee
where 
${\bf q}=n{\bf e_1}+m{\bf e_2}$, 
${\bf e_1} =\frac{4\pi}{\sqrt{3}\Lambda_b}\hat y$, 
${\bf e_2} =\frac{2\pi}{\Lambda_b}\hat
x-\frac{2\pi}{\sqrt{3}\Lambda_b}\hat y$.
This gives the cohesive energy
\be 
E_{coh}=\frac{4\pi}{\sqrt{3}l_0^2\Lambda^2_b}\sum_{\bf q}
\tilde U_{HF}(q)\left(\frac{R}{A l_0^2 q}\,J_1(qR)\right)^2.
\ee

With our effective interaction the sums in stripe and bubble
calculations 
converge quickly as we increase the cut off for ${\bf q}$. 
Ref.~\cite{Fogler}, which deals with electrons in the higher
electronic LLs,   
finds that the transition from the bubble to the stripe phase 
occurs at $\nu_n\approx 0.4$, where $\nu_n$ is the electron
filling in the $(n+1)^{st}$ 
Landau level.  Somewhat surprisingly, we also find a 
transition at $\nu_n^*\approx 0.4$ in all of our calculations.

As far as the comparison of stripe and bubble phases is
concerned, it is sufficient 
to know the cohesive energy; 
the energy of the reference, uniform HF state need not be
evaluated.   
However, to compare with the Fermi sea, the FQHE state, or the
paired state, 
we also need to know the energy of the uniform HF state, which
must be subtracted from the 
full energy to obtain the cohesive energy.

To calculate the energy of the uniform HF state we assume that
the CF-CF, 
background-background, and CF-background interactions are all of
the 
same form, characterized by $V^{CF}_m$. Since the direct
interaction terms cancel 
each other the only remaining part is the exchange term of the
CF-CF interaction.
Thus the energy of the uniform state $E_0$ is  
\be E_{0}=-\frac{U(0)}{2}\times\nu\ee
For example in the case of the Coulomb interaction:
\be E_{0}=-\frac{U(0)}{2}\times\nu=-\frac{1}{2}\int_0^{\infty} 
dq\,e^{\frac{1}{2}q^2}\times\nu=-\sqrt{\frac{\pi}{8}}\times\nu
\ee 
which is a familiar result.  For other interactions, 
we evaluate $E_0$ numerically.

\section{Results and Discussion}

\begin{figure}
\centerline{\psfig{figure=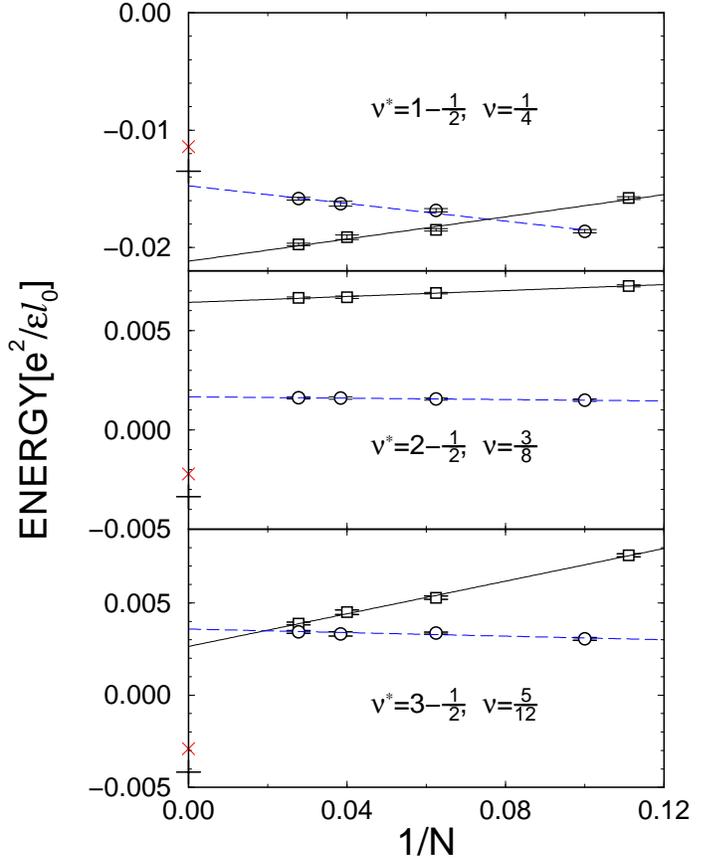,width=4.0in,angle=0}}
\caption{The cohesive energy per particle for the CF Fermi sea
(squares)
and the CF paired state (circles) as a function of $N$, the
number of CF ``holes" in the
topmost CF level.  The thermodynamic energies are also shown for
the CF
stripe and bubble phases by a dash ($-$) and cross ($\times$) on
the y-axis, respectively.  All energies are
measured relative to the uncorrelated uniform density state,
explained in the text.
The corresponding results for CF ``particles" at $\nu^*=1+1/2$,
$\nu^*=2+1/2$, and $\nu^*=3+1/2$,
corresponding to $\nu=3/8$, 5/12, and 7/16, were given in
Ref.~\protect\cite{Seung}. 
} 
\label{g4}
\end{figure}

\begin{figure}
\centerline{\psfig{figure=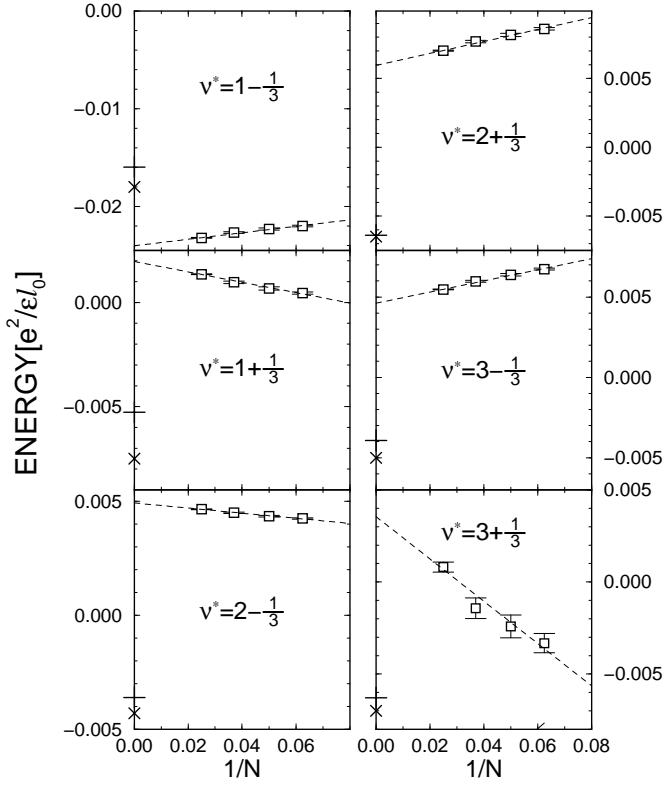,width=3.5in,angle=0}}
\caption{The cohesive energy per particle for composite fermions
at $\nu^*=n\pm 1/3$, which 
correspond to electron filling factors $\nu=\frac{3n\pm
1}{6n+3\pm 2}$, for the FQHE state
as a function of $N$, the number of composite fermions in the
partially filled composite
fermion Landau level.  This energy is shown by squares.  The
dash ($-$) and cross ($\times$) on the y-axis 
show the cohesive energies of the stripe and bubble phases,
respectively.} 
\label{g5}
\end{figure}

The energies of various states are shown in Fig.~2 of
Ref.~\onlinecite{Seung}, 
and Figs.~(\ref{g4}) and (\ref{g5}) of the present manuscript.
The lowest energy for the stripe or bubble states are determined
by varying the period.
We draw the following conclusions (subject to the validity of
the model):

(i) {\bf FQHE}: Within the lowest $^2$CF Landau level, CF holes
capture two more 
vortices to become $^4$CFs.  They show quantum Hall effect at 
$\nu_n^*=1/3$, which corresponds to $^2$CF filling factor of
$\nu^*=1-\frac{1}{3}$, and 
electron filling factor of $\nu=\frac{2}{7}$.
At $\nu_n^*=1/2$, which corresponds to $\nu=1/4$, the Fermi sea
has the lowest energy.
These results are consistent with the observations of FQHE at
2/7 and CF Fermi sea at 1/4.

(ii) {\bf Stripes:} At $\nu^*=n+\frac{1}{2}$, which correspond
to $\nu=(2n+1)/[2(2n+2)]$,
the stripe phase has the lowest energy (for $n\geq 1$).
The stripe phase is obtained independently of whether we model
$\nu^*=n+\frac{1}{2}$ 
as $1/2$ filling of CF particles on the background of $n$ filled
$^2$CF Landau levels, or 
as $1/2$ filling of CF holes on the background of $n+1$ filled
$^2$CF Landau levels, which 
suggests that the result is robust.
It should be noted, however, that the issues regarding the
stability of the unidirectional
charge density wave against a modulation along the length and
also against quantum fluctuations,
which have been investigated in the context of electronic
stripes in higher Landau
levels \cite{Fradkin}, have not been considered here.

The lowest energy for stripes 
is obtained for period $\Lambda/l_0=$ 10, 28, and 34 for
$\nu=3/8$, 5/12, and 7/16.
The period is rather large compared to that for the electron
stripes in higher LLs
(for which $\Lambda/l_0$ is of order unity), which is not
surprising because
the interaction between composite fermions is rather weak, and
also
the difference between the densities of the FQHE states on
either side is quite small.

A transport anisotropy in higher electronic Landau levels,
interpreted in terms of stripe
formation, is observed at temperatures below $\sim $ 50 mK
\cite{anisotropy}.  The
conditions for the CF stripes are more stringent.  Estimates of
the
critical temperature from the Hartree Fock theory are not
quantitatively reliable, but noting that
the effective interaction between composite fermions at
$\nu^*=n+1/2$ is
roughly an order of magnitude smaller than for electrons at
$\nu=n+1/2$,
as measured by the pseudopotentials,
we expect the critical temperature to also be similarly reduced.
Also, the much larger period suggests the need for a high degree
of density homogeneity.

(iii) {\bf Bubbles:} 
At $\nu^*=n\pm\frac{1}{3}$ with $n\geq 1$, the bubble phase has
the lowest energy. 
We remind the reader, however, that our trial wave functions for
FQHE states that 
work very well in the 
lowest Landau level are not so good in higher Landau levels,
which makes our calculation 
somewhat biased against FQHE.

\begin{figure}
\centerline{\psfig{figure=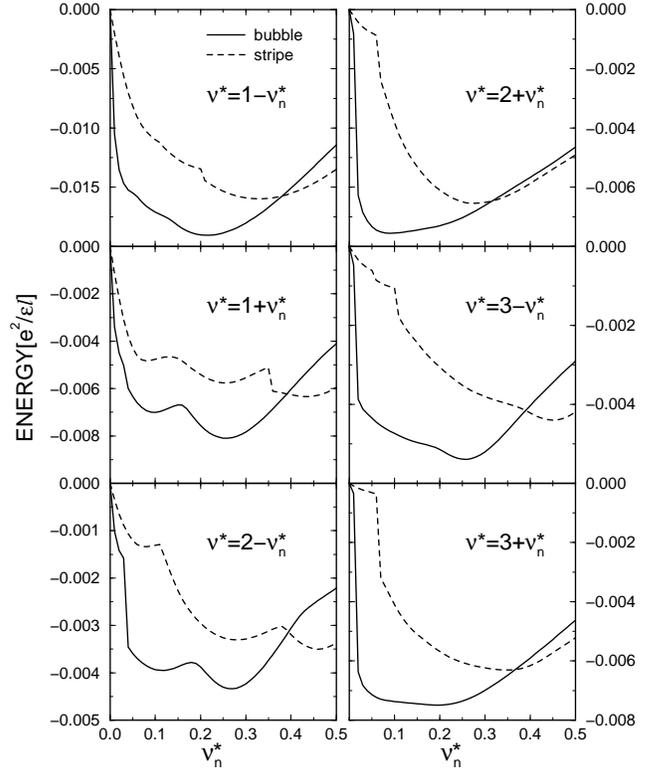,width=3.5in,angle=0}}
\caption{The cohesive energies of stripes (dashed line) and
bubbles (solid line) as a function 
of the CF filling factor in various CF Landau levels. The
striped state has lower energy in the
vicinity of $\nu^*=n+1/2$.} 
\label{g7}
\end{figure}

\begin{figure}
\centerline{\psfig{figure=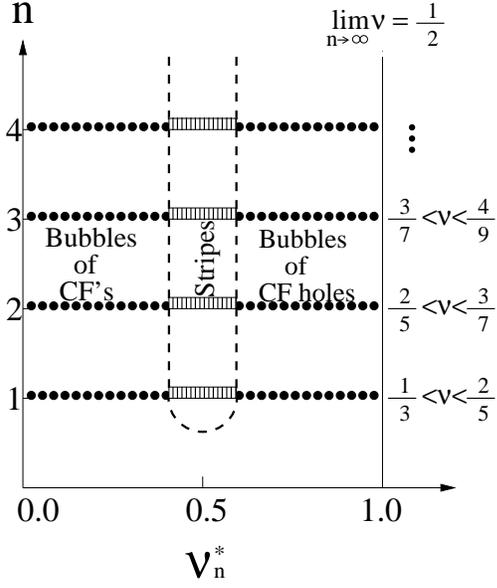,width=4.0in,angle=0}}
\caption{Phase diagram of various CF states, with 
 bubbles (solid dots), CF stripes (shaded region) and CF FQHE
(solid
line) as a function of the CF filling $\nu^*=n+\nu_n^*$.} 
\label{g10}
\end{figure}

To estimate at what filling a transition occurs from the bubble
crystal to the 
stripe phase, we have determined their energies as a function of
the filling factor, shown in
Fig.~(\ref{g7}).  The stripes are found to be stable
approximately in the region $0.4<\nu^*<0.6$,
outside of which bubbles take over. 
Overall, the phase diagram for various states of composite
fermions, 
shown schematically in Fig.~\ref{g10}, 
is remarkably similar to that for electrons.  For electrons: (i)
FQHE occurs in the 
lowest Landau level; (ii) stripes are believed to be relevant in
the vicinity of 
$\nu=n+1/2$ for $n\geq 2$; (iii) bubble or Wigner crystal takes
over in higher Landau 
levels for $\nu=n+\nu'$ with $\nu'<0.4$.  The behavior for
$^2$CFs is quite analogous.

\section{Finite thickness}

\begin{figure}
\centerline{\psfig{figure=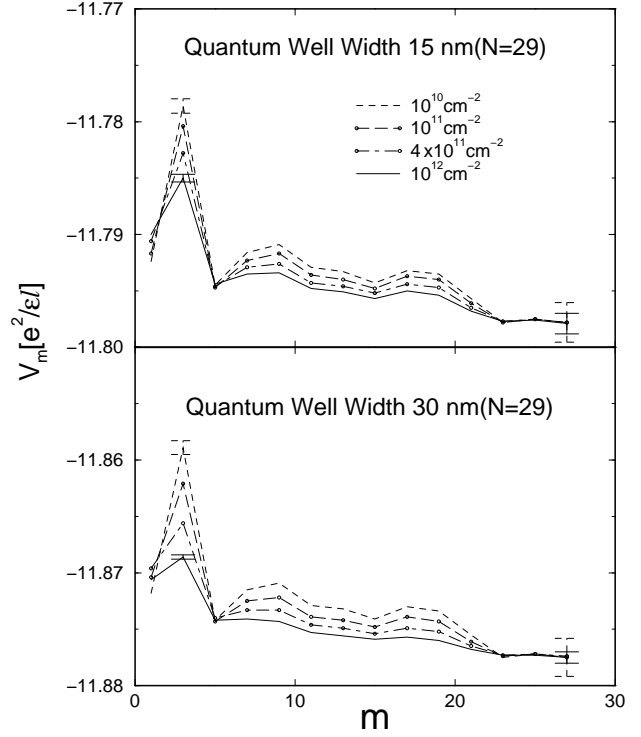,width=4.0in,angle=0}}
\caption{Pseudopotentials for the CF-CF interaction in the 2nd
CF-LL for various densities for 
quantum wells of widths 15 and 30 nm.  The inter-electron
interaction has been 
obtained in a local-density approximation.} 
\label{g8}
\end{figure}

\begin{figure}
\centerline{\psfig{figure=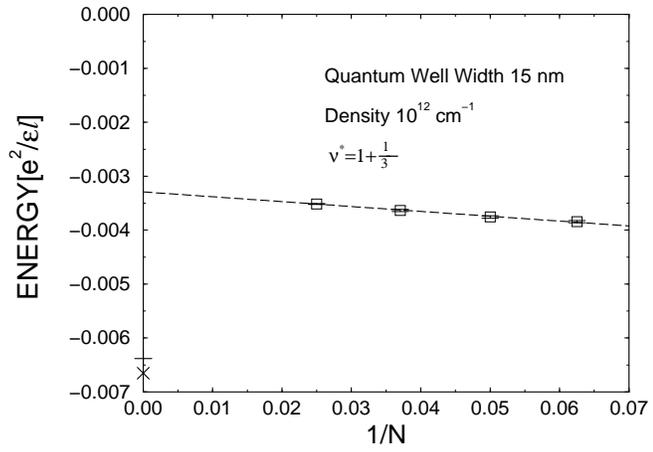,width=4.0in,angle=-90}}
\caption{Same as in Fig.~(\ref{g5}) for the modified
interaction.} 
\label{g9}
\end{figure}

We have assumed until now that the electron layer width is zero.  
In actual experiments, the electron wave function has a finite
extent in the 
transverse direction, which modifies the interaction
pseudopotentials. 
The modified interaction has been obtained in a local-density
approximation 
\cite{DasSarma,Meskini}.
To estimate how finite thickness affects the results presented
above, 
we have calculated the pseudopotentials for composite fermions
in the second 
CF-LL and found (Fig.~\ref{g8}) that the value $V^{CF}_3$, which
is the largest 
pseudopotential,
is reduced as we increase the density of the electrons. However,
the change is 
not large enough to alter the previous results.  As seen in
Fig.~(\ref{g9}),  
at $\nu=4/11$, where the composite fermion filling is
$\nu^*=4/3$, 
the bubble crystal phase continues to be most favorable among
the ones 
studied.

\section{Conclusion}

In summary, we have considered theoretically the question of
what new states of 
composite fermions are feasible as a result of the residual
interaction 
between composite fermions.  For this purpose, we have
constructed a model for the CF-CF
interaction, and studied various plausible states within a
variational scheme.
Our results suggest that fractional quantum Hall effect,  
stripe phase, as well as bubble crystal of composite fermions
can all occur at various 
filling factors.

\begin{center}
{\bf Acknowledgements}
\end{center}

This work was supported in part by the National Science
Foundation under grants no.
DMR-9986806 and DGE-9987589.  We are grateful to the High
Performance  
Computing (HPC) Group led by V.
Agarwala, J. Holmes, and J. Nucciarone, at the Penn State
University ASET (Academic Services
and Emerging Technologies) for
assistance and computing time with the LION-XE cluster.

\begin{center}{\bf Appendix}\end{center}

One of the most time consuming aspects of our Monte Carlo
calculation is the evaluation 
of determinants.  
When two matrices have many common columns (rows), the
determinant of one can be
related to the other, which amounts to substantial saving of
computation time in Monte 
Carlo.  Here we describe a general method for it.

Consider two $n\times n$ matrices ${\bf A}$ and ${\bf A'}$,
which 
differ only in the ($i_1,i_2,\cdots,i_m$)th columns, where $0<
m\leq n$.
Our aim is to obtain the determinant of ${\bf A'}$ with minimum
computation.
The determinant of ${\bf A'}$ can be written as 
\be 
\det[{\bf A'}]=\det[{\bf A}]\det[{\bf A^{-1}A'}]
\ee
Since $A'$ differs from $A$ only at ($i_1,i_2,\cdots,i_m$)th
columns, 
$[{\bf A^{-1}A'}]_{k l}$ is equal to $\delta_{k l}$ when
$l\notin (i_1,i_2,\cdots,i_m)$

Define ${\bf C}_j$ ($j=1,2,\cdots,m$) to be $n\times n$ matrices 
whose components are same as those of ${\bf A}$ at $j$th column
and zero elsewhere.
Further, define an $m\times m$ matrix ${\bf B}$ such that 
$[{\bf B}]_{a b}=[{\bf A^{-1}C'_b}]_{i_a i_b}$ wher
e $a,b\in\{1,2,\cdots,m\}$. Then the following is true.
\be \det[{\bf A^{-1}A'}]=\det[{\bf B}] \ee

This gives a simple relation between the determinants of ${\bf
A}'$ and  
${\bf A}$: 
\bea\det[{\bf A'}]=\det[{\bf A}]\det[{\bf B}]
\eea

Consider some examples:

1. Updating of one column:\\
\be \det[{\bf A'}]=\det[{\bf A}]\det[{\bf B}]=\det[{\bf A}][{\bf
A^{-1}C'_{1}}]_{i_1 i_1}\ee
2. Updating of two columns:\\
\bea
\det[{\bf A'}]&=&\det[{\bf A}]\det[{\bf B}] \nonumber \\
&=&\det[{\bf A}]\big([{\bf A^{-1}C'_{1}}]_{i_1 i_1}[{\bf
A^{-1}C'_{2}}]_{i_2 i_2} \nonumber \\
&&~~~~~~~-[{\bf A^{-1}C'_{1}}]_{i_2 i_1}[{\bf
A^{-1}C'_{2}}]_{i_1 i_2}\big)\eea
3. Updating of three columns:
\bea
\det[{\bf A'}]&=&\det[{\bf A}]\det[{\bf B}] \nonumber \\
&=&\det[{\bf A}]\times \nonumber \\
&& \big([{\bf A^{-1}C'_{1}}]_{i_1 i_1}
[{\bf A^{-1}C'_{2}}]_{i_2 i_2}[{\bf A^{-1}C'_{3}}]_{i_3 i_3}
\nonumber \\
&&+[{\bf A^{-1}C'_{1}}]_{i_2 i_1}
[{\bf A^{-1}C'_{2}}]_{i_3 i_2}[{\bf A^{-1}C'_{3}}]_{i_1 i_3}
\nonumber \\
&&+[{\bf A^{-1}C'_{1}}]_{i_3 i_1}
[{\bf A^{-1}C'_{2}}]_{i_1 i_2}[{\bf A^{-1}C'_{3}}]_{i_2 i_3}
\nonumber \\
&&[{\bf A^{-1}C'_{1}}]_{i_2 i_1}
[{\bf A^{-1}C'_{2}}]_{i_1 i_2}[{\bf A^{-1}C'_{3}}]_{i_3 i_3}
\nonumber \\
&&-[{\bf A^{-1}C'_{1}}]_{i_3 i_1}
[{\bf A^{-1}C'_{2}}]_{i_2 i_2}[{\bf A^{-1}C'_{3}}]_{i_1 i_3}
\nonumber \\
&&-[{\bf A^{-1}C'_{1}}]_{i_1 i_1}
[{\bf A^{-1}C'_{2}}]_{i_3 i_2}[{\bf A^{-1}C'_{3}}]_{i_2
i_3}\big)
\eea

\pagebreak

\begin{table}
\caption{Pseudopotentials for CF-holes in the lowest CF Landau
level. In this and the following 
tables, $E_m$ is the energy of the 
full state, and $\Delta E_m$ is the Monte Carlo statistical
uncertainty.
The energies are quoted in units of $e^2/\epsilon l_0$.
\label{tab1}}
\begin{tabular}{ccc}
\hline
$m$ & $E_m$      & $\Delta E_m$         \\
\hline
1   &-7.726983   &0.0005744120          \\              
3   &-7.757470   &0.0007778533          \\
5   &-7.752651   &0.0010310604          \\
7   &-7.758716   &0.0010988030          \\
9   &-7.758750   &0.0012034221          \\      
11  & -7.759422  &0.0012985310          \\
13  & -7.760764  &0.0012624134          \\
15  & -7.761201  &0.0012929297          \\
17  & -7.762082  &0.0016546843          \\      
19  & -7.762933  &0.0016976433  
\end{tabular}
\end{table}

\begin{table}
\caption{Pseudopotentials for CF-particles in the 2nd CF Landau
level.
\label{tab2}}
\begin{tabular}{ccc}
\hline
$m$ & $E_m$      & $\Delta E_m$     \\
\hline
1   & -11.755883 & 0.0005593626     \\
3   & -11.741289 & 0.0006092897    \\
5   & -11.756699 & 0.0005888317    \\
7   & -11.754222 & 0.0005821469   \\
9   & -11.754109 & 0.0006913264   \\
11  & -11.755223 & 0.0006993560   \\
13  & -11.755586 & 0.0008570091   \\
15  & -11.755554 & 0.0005545685   \\
17  & -11.756373 & 0.0006021690   \\
19  & -11.756839 & 0.0005796026   \\
21  & -11.756438 & 0.0005914016 \\
23  & -11.756528 & 0.0006283216 \\
25  & -11.756233 & 0.0006583812 \\
27  & -11.756788 & 0.0006599042
\end{tabular}
\end{table}

\begin{table}
\caption{Pseudopotentials for CF-holes in the 2nd CF Landau
level.
\label{tab3}}
\begin{tabular}{ccc}
\hline
$m$ & $E_m$      & $\Delta E_m$     \\
\hline
1   &-18.135544  & 0.0009774497  \\
3   &-18.120577  & 0.0009665822 \\
5   &-18.133069  & 0.0009376358 \\
7   &-18.131464  & 0.0008402889 \\
9   &-18.130383  & 0.0009074839 \\
11  & -18.130536 & 0.0009764340 \\
13  & -18.131314 & 0.0009651683 \\
15  & -18.130754 & 0.0010156039 \\
17  & -18.131685 & 0.0009203913 \\
19  & -18.130155 & 0.0009442089 \\
21  & -18.131228 & 0.0010169967
\end{tabular}
\end{table}

\begin{table}
\caption{Pseudopotentials for CF-particles in the 3rd CF Landau
level.
\label{tab4}}
\begin{tabular}{ccc}
\hline
$m$ & $E_m$      & $\Delta E_m$     \\
\hline
1   & -18.942399 & 0.0010950836     \\
3   & -18.932547 & 0.0012154270     \\
5   & -18.927645 & 0.0011055870     \\
7   & -18.935310 & 0.0012656451     \\
9   & -18.934032 & 0.001233024      \\
11  & -18.934417 & 0.0012795043     \\
13  & -18.933864 & 0.0013178322     \\
15  & -18.933060 & 0.0012459279     \\
17  & -18.935017 & 0.0013152929     \\
19  & -18.934135 & 0.0013808266     \\
21  & -18.934547 & 0.0015265486     \\
23  & -18.934336 & 0.0012488644     
\end{tabular}          
\end{table}

\begin{table}
\caption{Pseudopotentials for CF-holes in the 3rd CF Landau
level.
\label{tab5}}
\begin{tabular}{ccc}
\hline
$m$ & $E_m$      & $\Delta E_m$     \\
\hline
1   & -17.620082 & 0.000983051 \\
3   & -17.612904 & 0.000917553  \\
5   & -17.606845 &  0.001063009 \\
7   & -17.614824 &  0.001009564 \\
9   & -17.615345 &  0.000978099 \\
11  & -17.614743 &  0.000939712 \\
13  & -17.614587 & 0.000958717  \\
15  & -17.612469 &  0.000891311 
\end{tabular}
\end{table}

\begin{table}
\caption{Pseudopotentials for CF-particles in the 4th CF Landau
level.
\label{tab6}}
\begin{tabular}{ccc}
\hline
$m$ & $E_m$      & $\Delta E_m$     \\
\hline
1   &-16.691824    &0.00101485  \\
3   &-16.686846    &0.00105839  \\
5   &-16.682793    &0.00107136 \\
7   &-16.680260    & 0.00108352 \\
9   &-16.687565    &0.00107323  \\
11  &-16.684489    & 0.00111337  \\
13  &-16.685492    & 0.00125616  \\
15  &-16.685188    &  0.00095845 
\end{tabular}
\end{table}

\begin{table}
\caption{Parameters for our model interaction for two CF
particles in an otherwise 
empty CF-LL.  The second, third, and fourth CF-LLs are
considered. All of the $c_i$'s not shown
here are set equal to zero.
\label{b1}}
\begin{tabular}{|c|c|c|c|}
parameter & 2nd CF-LL & 3rd CF-LL & 4th CF-LL \\
\hline
$c_2$  & -1.57357 & 1.75418 & -10.6790\\
$c_6$  & 0.00817294 & -0.0114760 & 0.0793665 \\
$c_{10}$  & -1.72040E-6  & 2.62965E-6 & -2.49751E-5 \\
$c_{14}$  & 4.31056E-11 & -6.30827E-11 & 9.87564E-10 \\
$c_{18}$  & -1.80465E-16 & 1.96255E-16 & -7.12056E-15 \\
$c_{22}$  & 0.0 & 0.0 & 9.29168E-21 
\end{tabular}
\end{table}

\begin{table}
\caption{Parameters for our model interaction for two CF holes
in an otherwise 
full CF-LL.  The first, second and third CF-LLs are considered.
All of the $c_i$'s not shown
here are set equal to zero.
\label{b2}}
\begin{tabular}{|c|c|c|c|}
parameter & 1st CF-LL & 2nd CF-LL & 3rd CF-LL \\
\hline
$c_2$ & 2.08604 & -0.641674 & 1.15620 \\ 
$c_6$ & -0.0122001 & 0.00240019 & -0.00699306 \\
$c_{10}$& 2.85410E-6 & -2.15587E-7 & 1.38513E-6\\ 
$c_{14}$& -7.63371E-11 & -1.27142E-12 & -2.49865E-11 \\
$c_{18}$& 2.90620E-16 & 1.48118E-17 & 4.72082E-17  
\end{tabular}
\end{table}

\end{document}